\documentclass[twocolumn,showpacs,preprintnumbers,amsmath,amssymb,APSl,prd,superscriptaddress]{revtex4-1}

\usepackage{dcolumn}
\usepackage{bm}
\usepackage{ifpdf}
\usepackage{hyperref}
\usepackage{dcolumn}
\usepackage{bm}
\usepackage[spanish,english]{babel}
\usepackage{amsfonts}
\usepackage{amssymb}
\usepackage{graphicx}
\usepackage[latin1]{inputenc}
\newcommand{\R}{\mathcal{R}}

\newcommand{\be}{\begin{equation}}
\newcommand{\en}{\end{equation}}
\newcommand{\bea}{\begin{eqnarray}}
\newcommand{\ena}{\end{eqnarray}}

\begin{document}

\title{Black hole formation from a null fluid in extended Palatini gravity}

\author{Jesus Martinez-Asencio}
\author{Gonzalo J. Olmo} \email{ gonzalo.olmo@csic.es}
\affiliation{Departamento de F\'{i}sica Te\'{o}rica and IFIC, Centro Mixto Universidad de
Valencia - CSIC. Universidad de Valencia, Burjassot-46100, Valencia, Spain}
\author{D. Rubiera-Garcia} \email{rubieradiego@gmail.com}
\affiliation{Departamento de F\'{i}sica, Universidad de Oviedo, Avenida Calvo Sotelo 18, 33007, Oviedo, Asturias, Spain}


\date{\today}

\begin{abstract}
We study the formation and perturbation of black holes by null fluxes of neutral matter in a quadratic extension of General Relativity formulated \`{a} la Palatini. Working in a spherically symmetric space-time, we obtain an exact analytical solution for the metric that extends the usual Vaidya-type solution to this type of theories. We find that the resulting space-time is formally that of a Reissner-Nordstr\"om black hole but with an effective charge term carrying the wrong sign in front of it. This effective charge is directly related to the luminosity function of the radiation stream. When the ingoing flux vanishes, the charge term disappears and the space-time relaxes to that of a Schwarzschild black hole. We provide two examples that illustrate the formation of a black hole from Minkowski space and the perturbation by a finite pulse of radiation of an existing Schwarzschild black hole.
\end{abstract}

\maketitle

\section{Introduction} \label{sec:I}

The properties and structure of black holes in extended theories of gravity have received increasing attention in the last years. In the context of $f(R)$ theories, in both their metric \cite{Capozziello2011,f(R)}  and Palatini formulations \cite{Olmo:2011uz} (see also \cite{Hybrid}), important progress has been made in the understanding of their thermodynamical properties \cite{faraoni}, on the existence of spherically symmetric \cite{Capo2012,Olmo:2011ja} and rotating solutions \cite{Dobado}, and also on the propagation of perturbations on such backgrounds \cite{AdlC+chinos}.

A natural motivation to explore the properties of black holes in extensions of General Relativity (GR) is to test the effects that new physics could have on these extreme scenarios. However, up to date the success within this approach has been quite limited because, very often, only constant curvature solutions are analytically accessible and yield solutions which are essentially equivalent to the ones found in GR with a cosmological constant (see, however, \cite{CB2005}). To see this, consider the field equations of $f(R)$ theories in the metric formulation,
\begin{equation}\label{eq:Metric}
f_R R_{\mu\nu}-\frac{f}{2}g_{\mu\nu}-\nabla_\mu\nabla_\nu f_R+g_{\mu\nu}\Box f_R=\kappa^2T_{\mu\nu} \ ,
\end{equation}
where we denote $\kappa^2\equiv 8\pi G$ and $f_R\equiv \partial f/\partial R$. From this equation, one finds that the Ricci scalar satisfies
\begin{equation}\label{eq:traceM}
3\Box f_R  +Rf_R-2f=\kappa^2T \ .
\end{equation}
This equation shows that in vacuum or with traceless sources such as electromagnetic fields,  $R=R_0=2f(R_0)/f_R(R_0)=$ constant solutions are possible. Such solutions turn (\ref{eq:Metric}) into
\begin{equation}\label{eq:GR+L}
R_{\mu\nu}=\frac{\kappa^2}{f_R(R_0)}T_{\mu\nu}+\frac{R_0}{4}g_{\mu\nu} \ ,
\end{equation}
where the last term plays the role of an effective cosmological constant, $4\Lambda_{eff}=R_0$, and $\kappa^2/f_R(R_0)$ simply rescales the relation between the bare and the observable Newton's constant, $G_{obs}=G/f_R(R_0)$. In the Palatini formulation, where metric and connection are treated as independent physical entities, the (metric) field equations take the form
\begin{eqnarray}\label{eq:Palatini}
R_{\mu\nu}-\frac{R}{2}g_{\mu\nu}&=&\frac{\kappa^2T_{\mu\nu} }{f_{\R}}-\frac{\R f_{\R}-f}{2f_{\R}}g_{\mu\nu} \nonumber \\ &-& \frac{3}{2f_{\R}^2}\Big[\partial_\mu f_{\R} \partial_\nu f_{\R}-\frac{1}{2}g_{\mu\nu}(\partial f_{\R})^2\Big] \\ &+& \frac{1}{f_{\R}}\left(\nabla_\mu\nabla_\nu f_{\R}-g_{\mu\nu}\Box f_{\R}\right) \nonumber ,
\end{eqnarray}
where the Palatini curvature ${\R}$ is a function of the trace $T$ of the matter that satisfies the algebraic equation
\begin{equation}\label{eq:traceP}
{\R}f_{\R}-2f=\kappa^2 T \ .
\end{equation}
In vacuum or with traceless sources, this last equation yields a constant curvature ${\R}={R}_0$, and  (\ref{eq:Palatini}) exactly boils down to (\ref{eq:GR+L}).

This degeneracy of the two versions of $f(R)$ theories with GR plus cosmological constant precludes any attempt to extract new physics by considering $f(R)$ extensions of GR in scenarios with constant curvature regardless of the symmetries of the problem. This, in particular, implies that also in the dynamical scenario of a null fluid collapsing to form a black hole, $f(R)$ theories yield the same result as in GR, which has been verified recently in the metric approach \cite{Ghosh:2012zz}.

Deviations from the predictions of GR in black hole space-times have been observed in the Palatini formulation of $f(\R)$ theories coupled to Born-Infeld non-linear electrodynamics \cite{Olmo:2011ja} and also in Ricci-squared extensions of Palatini theories coupled to Maxwell's electromagnetism \cite{Olmo:2012nx}.  In the former case, it was found that for particular choices of the Lagrangian, $\tilde{f}(\R)=R+a l_P^2 R^2$ with $l_P \equiv \sqrt{\hbar G/c^3}$ representing the Planck length, the intensity of the central singularity can be reduced as compared to GR, while in the latter non-singular charged black hole solutions were obtained in exact analytical form. The emergence of new physics in the Palatini $f(\R)$ case can be traced back to the fact that for a non-linear theory of electrodynamics the trace of the stress-energy tensor in four dimensions is not zero, which implies a non-trivial relation between $\R$ and $T$ via (\ref{eq:traceP}). In the extended Ricci-squared theory, the family of Lagrangians chosen, of the form $f(\R,Q)=\tilde{f}(\R)+l_P^2 \R_{\mu\nu} \R^{\mu\nu}$, implies that $\R=$ constant for the electromagnetic field but $Q=\R_{\mu\nu} \R^{\mu\nu}$ turns out to be non-constant. This contribution is crucial to change the inner structure of the black hole, which develops a central core whose area grows linearly with the charge in multiples of an elementary unit of order the Planck area. For a certain charge-to-mass ratio, which sets the core mass density precisely at the Planck scale, the space-time is completely regular. For any other choice of the charge-to-mass ratio, the core surface is singular (null-like singularity). In both the singular and non-singular cases, the metric quickly tends to that predicted by GR as one moves away from the core surface. This puts forward the fact that the emergence of the core and the removal of the singularity are due to non-perturbative mechanisms.  A similar qualitative behavior also appears for these models in the cosmological setting \cite{Barragan2010}, where the big bang singularity is replaced by a cosmic bounce at the Planck scale.

The non-perturbative character of the dynamics of Palatini $f(\R,Q)$ theories near the core, makes it difficult to foresee if the non-singular solutions found in \cite{Olmo:2012nx} are robust under small perturbations or not. If a perturbative treatment were possible, under small external perturbations one would expect a small response of the parameters of the system, thus making the non-singular solutions highly unstable. However, since there is evidence supporting a non-perturbative behavior, nearby configurations could dynamically tend to settle to the static regular solution by radiating away the excess of charge and mass. A proper answer to this question, therefore, can only be obtained by considering the full dynamical problem. In this paper we begin the analysis of this question and consider the simplest non-trivial dynamical scenario, namely, the formation and perturbation of spherically symmetric  black holes by null fluxes of neutral matter in the $f(\R,Q)$ Palatini model considered in \cite{Olmo:2012nx}. Due to the complicated new physics involved in the charged problem, in this work we focus on the case of Schwarzschild black holes and leave the analysis of the charged case for an independent publication. Here we will see that though the null fluid yields vanishing $\R$ and $Q$, modifications with respect to GR are manifest, even though the unperturbed (Schwarzschild solution) is the same as in GR, and can be computed analytically. In fact, the usual Vaidya-type metric found in GR, namely, $ds^2=-(1-2M(v)/r)dv^2+2dv dr+r^2d\Omega^2$ \cite{Vaidya70}, is now modified by a charge-like term with the wrong sign in front of it, i.e.,  $(1-2M(v)/r)\to (1-2M(v)/r-Q^2(v)/r^2)$, where $Q^2(v)$ is related to the {\it luminosity } of the radiation flux (the notation used here will be explained later). This result puts forward that in Palatini theories the stress-energy density of the matter fields, not just its integral, has a direct influence on the form of the space-time metric. In particular, this allows to see how a flux of radiation modifies the form of the space-time metric along its path and complements with an exact and non-perturbative solution the discussion presented in \cite{Olmo:2011sw}. The existence of this charge-like (transient) contribution opens a new window to the recovery of the classical and quantum information stored in the black hole via Hawking radiation. This point will also be discussed in this work.

The content of this paper is organized as follows. In section \ref{sec:II} we describe the main elements of Palatini formalism, work out the field equations for the metric and the connection, and finally express the relevant field equations using an auxiliary metric associated to the independent connection.  In section \ref{sec:III} we compute the relation between the auxiliary and physical metrics for a null fluid, and write explicitly the field equations for this problem. In section \ref{sec:IV} we consider the problem of formation and perturbation of black holes in this framework and illustrate it with two simple examples of luminosity functions. We conclude in section \ref{sec:V} with a summary and discussion of some future perspectives.

\section{Palatini $f(R,Q)$ theories} \label{sec:II}

We consider Palatini $f(R,Q)$ theories defined as follows

\be\label{eq:action}
S[g,\Gamma,\psi_m]=\frac{1}{2\kappa^2}\int d^4x \sqrt{-g}f(R,Q) +S_m[g,\psi_m],
\en
where $\kappa^2\equiv 8\pi G$,  $S_m[g,\psi_m]$ represents the matter action, $g_{\alpha\beta}$ is the space-time metric, $R=g^{\mu\nu}R_{\mu\nu}$, $Q=g^{\mu\alpha}g^{\nu\beta}R_{\mu\nu}R_{\alpha\beta}$, $R_{\mu\nu}={R^\rho}_{\mu\rho\nu}$, and ${R^\alpha}_{\beta\mu\nu}=\partial_{\mu}
\Gamma^{\alpha}_{\nu\beta}-\partial_{\nu}
\Gamma^{\alpha}_{\mu\beta}+ \Gamma^{\alpha}_{\mu\lambda}\Gamma^{\lambda}_{\nu\beta}-\Gamma^{\alpha}_{\nu\lambda}\Gamma^{\lambda}_{\mu\beta}$. The connection $\Gamma^{\alpha}_{\beta \gamma}$ has no a priori relation with the metric (Palatini formalism) and must be determined by the theory through the corresponding field equations.
Variation of (\ref{eq:action}) with respect to metric and connection leads to
\bea
f_R R_{\mu\nu}-\frac{f}{2}g_{\mu\nu}+2f_QR_{\mu\alpha}{R^\alpha}_\nu &=& \kappa^2 T_{\mu\nu}\label{eq:met-varX}\\
\nabla_{\beta}\left[\sqrt{-g}\left(f_R g^{\mu\nu}+2f_Q R^{\mu\nu}\right)\right]&=&0.  \label{eq:con-varX}
\ena
Here $T_{\mu\nu}$ is the stress-energy tensor of the matter sector $S_m$, obtained as $T_{\mu\nu}=-\frac{2}{\sqrt{-g}}\frac{\delta S_m}{\delta g^{\mu\nu}}$. In the above derivation of the field equations two assumptions have been made, namely, vanishing torsion, $\Gamma_{[\beta \gamma]}^{\alpha}=0$, and symmetric Ricci tensor, $R_{[\mu\nu]}=0$ (see \cite{Olmo12z} for the resulting equations and implications when such constraints are relaxed). In order to solve for the connection, we must note that Eq.(\ref{eq:con-varX}) depends on the metric $g_{\mu\nu}$ and on the independent connection $\Gamma^\alpha_{\beta\gamma}$, though the connection dependence can be eliminated in favor of the matter. This can be seen by defining a matrix $\hat{P}$ whose components are ${P_{\alpha}}^{\nu}=R_{\alpha \beta}g^{\beta \nu}$, and rewriting (\ref{eq:met-varX}) as

\be
2f_Q\hat{P}^2+f_R \hat{P}-\frac{f}{2}\hat{I} = \kappa^2 \hat{T} \label{eq:met-varRQ2} \ ,
\en
where $\hat{I}$ and $\hat{T}$ are the matrix representation of the identity $\delta_{\mu}^{\nu}$ and the stress-energy tensor ${T_{\mu}}^{\nu}$, respectively. Eq.(\ref{eq:met-varRQ2}) implies that $\hat{P}=\hat{P}(\hat{T})$ is a function of the matter and therefore $R={P_\alpha}^{\alpha}$ and $Q={P_\alpha}^{\nu} {P_\nu}^{\alpha}$ are also functions of the matter. To solve the equation for the independent connection (\ref{eq:con-varX}), we look for another (auxiliary) metric $h_{\mu\nu}$ such that $\Gamma^{\alpha}_{\beta \gamma}$ becomes its Levi-Civita connection. To do it so, we propose the following ansatz \cite{Olmo:2012nx} $\sqrt{-g}\left(f_R g^{\mu\nu}+2f_Q R^{\mu\nu}\right)=\sqrt{-h} h^{\mu\nu}$, which turns (\ref{eq:con-varX}) into $\nabla_{\beta}[\sqrt{-h}h^{\mu\nu}]=0$. Simple algebraic manipulations show that the relation between $g_{\mu\nu}$ and $h_{\mu\nu}$ is given by

\be \label{eq:h-g-relation}
\hat{h}^{-1}=\frac{\hat{g}^{-1}\hat\Sigma}{\sqrt{\det\hat\Sigma}} \ , \
\hat{h}=\left(\sqrt{\det\hat\Sigma}\right)\hat\Sigma^{-1}\hat{g} ,
\en
which shows that the connection of $f(R,Q)$ theories with the constraints above can be explicitly solved in terms of the physical metric $g_{\mu\nu}$ and the matter sources. In (\ref{eq:h-g-relation}) we have introduced the object
\be
{\Sigma_\alpha}^{\nu}=\left(f_R \delta_{\alpha}^{\nu} +2f_Q {P_\alpha}^{\nu}\right),
\en
which contains all the information about the relative deformation between $g_{\mu\nu}$ and $h_{\mu\nu}$. In particular, in the case of $f(R)$ theories we have ${\Sigma_\alpha}^{\nu}={f_R}\delta_{\alpha}^{\nu}$ and, therefore, $g_{\mu\nu}$ and $h_{\mu\nu}$ become conformally related as $g_{\mu\nu}=\frac{1}{f_R}h_{\mu\nu}$. In terms of the metric $h_{\mu\nu}$ the metric field equation (\ref{eq:met-varX}) can be written in a more compact and transparent form. To see this, note that (\ref{eq:met-varRQ2}) can be expressed as $\hat{P} \hat{\Sigma}=\frac{f}{2}\hat{I}+k^2 \hat{T}$ and taking into account that $\hat{P}\hat{\Sigma}=\sqrt{\det \hat{\Sigma}} R_{\mu\alpha}(h) h^{\alpha \nu}$ we arrive at
\be
{R_\mu}^\nu (h) =\frac{1}{\sqrt{\det \hat\Sigma}}\left(\frac{f}{2}{\delta_\mu}^\nu +\kappa^2 {T_\mu}^\nu\right) \label{eq:met-varRQ4}.
\en
This form of the field equations describes the dynamics of Palatini $f(R,Q)$ theories in such a way that the right-hand side only depends on the matter sources, as both $R$ and $Q$ (and hence $f$) are functions of the stress-energy tensor. It should be noted that, unlike in the usual metric formalism, $h_{\mu\nu}$ satisfies second-order equations and, therefore, this type of Palatini $f(R,Q)$ theory is ghost-free.

\section{Field equations for a null fluid} \label{sec:III}

In this section we consider a specific form of matter that will allow us to explicitly obtain the matrix ${\Sigma_\alpha}^{\nu}$, a key element of the field equations of our theory. This will allow us to express the field equations in a form suitable for calculations. Then we will specify the $f(R,Q)$ model and will write the field equations taking advantage of the simplifications offered by the assumed spherical symmetry.

\subsection{Determination of ${\Sigma_\alpha}^{\nu}$}

We shall consider the formation of black holes from an ingoing flux of presureless null neutral matter whose stress-energy tensor is written as
\be \label{eq:ingoingfluid}
T_{\mu\nu}=\rho_{in} l_{\mu}l_{\nu} \ ,
\en
where $\rho_{in}$ is the energy density of the ingoing stream and $l_\mu$ represents a null radial vector, $l_\mu l^\mu=0$. With this matter source, (\ref{eq:met-varRQ2}) can be rewritten as
\be
2f_Q\left(\hat{P}+\frac{f_R}{4f_Q} \hat{I}\right)^2=\left(\frac{f}{2}+\frac{f_R^2}{8f_Q} \right) \hat{I}+k^2 \hat{T}  \label{eq:fieldroot} \ ,
\en
where $\hat T\equiv  \rho_{in} l_{\mu}l^{\nu}$. Since to construct ${\Sigma_\alpha}^{\nu}$ we must have an expression for $\hat P$, we need to take the square root of the above equation. To proceed, we propose the following ansatz
\be \label{eq:Bdefinition}
{B_\mu}^{\alpha}=\lambda \delta_{\mu}^{\alpha} + \Omega_{in} l_{\mu}l^{\alpha},
\en
as the square root of the left-hand side of (\ref{eq:fieldroot}), with $\lambda$ and $\Omega_{in}$ two functions to be determined. We then find that
\be
{B_\mu}^{\alpha}{B_{\alpha}}^\nu=\lambda^2 \delta_{\mu}^{\nu}+2\lambda \Omega_{in} l_{\mu}l^{\nu}
\en
and comparing with the right-hand side of (\ref{eq:fieldroot}), we obtain
\be
\lambda^2=\frac{f}{2}+\frac{f_R^2}{8f_Q} \ , \ \Omega_{in}=\frac{k^2 \rho_{in}}{2\lambda} \ .
\en
We thus find that
\be \label{eq:sigmaparticular}
\hat{\Sigma}=\frac{f_R}{2}\hat{I}+\sqrt{2f_Q} \hat{B}=\left[\frac{f_R}{2}+\sqrt{2f_Q}\lambda \right]\delta_{\mu}^{\nu} + \sqrt{\frac{f_Q}{2}} \frac{\kappa^2 \rho_{in}}{\lambda}l_{\mu}l^{\nu}.
\en
This equation implies that once a Lagrangian $f(R,Q)$ is given, the relation between $g_{\mu\nu}$ and $h_{\mu\nu}$ is immediately obtained and, consequently, we will be able to write explicitly the solution in the metric $g_{\mu\nu}$ once its counterpart in the metric $h_{\mu\nu}$ is determined.

\subsection{$f(R,Q)$ model and field equations}

Therefore, to proceed further we must specify the $f(R,Q)$ model. In this work we will focus on the Lagrangian \cite{Olmo:2012nx,Barragan2010,OSAT2009}
\be \label{eq:Palatinimodel}
f(R,Q)=R+l_P^2 (aR^2 + Q),
\en
where $a$ is some dimensionless constant and $l_P$ is Planck's length. Lagrangians of this kind are expected to arise as effective descriptions or in the low-energy limit of quantum theories of gravity in which both metric and connection play relevant roles in the resulting effective geometry (see e.g. \cite{QFTCS,string,OS09}). For this model,  tracing in (\ref{eq:met-varX}) with $g^{\mu\nu}$ leads to $R=-k^2 T$, which is the same relation as in GR. Since for a null fluid $T=0$, we have $R=0$ and thus $f_R=1$ and $f=l_P^2 Q$. With these data, $Q$ can be calculated from the trace of (\ref{eq:Bdefinition}),
\be
\frac{1}{2 \sqrt{2}l_P}= \sqrt{\frac{l_P^2 Q}{2}+\frac{1}{8l_P^2}} \ ,
\en
which is satisfied only if $Q=0$. Using this result and the fact that $R=0$ we can finally put $\hat{\Sigma}$ as
\be \label{eq:sigmafinal}
{\Sigma_\mu}^\nu= \delta_{\mu}^{\nu} + 2l_P^2 \kappa^2 \rho_{in} l_{\mu}l^{\nu} \ .
\en
With some little algebra one can verify that $\det \hat{\Sigma}=1$. Collecting all these results we find that the field equations (\ref{eq:met-varRQ4}) boil down to
\be \label{eq:fieldequationsdefinitive}
{R_{\mu}}^{\nu}(h)=\kappa^2 \rho_{in} l_{\mu}l^{\nu} \ ,
\en
which can also be written using the Einstein tensor of the metric $h_{\mu\nu}$ as
\be \label{eq:fieldequationsdefinitiveX}
{G_{\mu}}^{\nu}(h)=\kappa^2 \rho_{in} l_{\mu}l^{\nu} \ .
\en
The formal simplicity of this expression puts forward the great advantage of using the metric $h_{\mu\nu}$ instead of the physical metric $g_{\mu\nu}$ to deal with the field equations. To see this, note that $\Gamma^\alpha_{\beta\gamma}$ (the connection that defines $R_{\mu\nu}(h)$) can be expressed in terms of $g_{\mu\nu}$ as
\begin{equation}
\Gamma^\alpha_{\beta\gamma}=L^\alpha_{\beta\gamma}+A^\alpha_{\beta\gamma} \ ,
\end{equation}
where $L^\alpha_{\beta\gamma}$ is the Levi-Civita connection of $g_{\mu\nu}$ and $A^\alpha_{\beta\gamma}$ is a tensor given by
\begin{equation}
A^\alpha_{\beta\gamma}=\frac{h^{\alpha\rho}}{2}\left[\nabla^L_\beta h_{\rho\gamma}+\nabla^L_\gamma h_{\rho\beta}-\nabla^L_\rho h_{\beta\gamma}\right] \ ,
\end{equation}
with $\nabla^L_\beta W_\nu=\partial_\beta W_\nu-L^\sigma_{\beta\nu}W_\sigma$. Expressing (\ref{eq:fieldequationsdefinitiveX}) in terms of $g_{\mu\nu}$ and the matter would lead to an expression of the form $G_{\mu\nu}(g)=\kappa^2 \rho_{in} l_{\mu}l_{\nu}+$terms involving up to second-order derivatives of $\rho_{in}$ and $l_\nu$. Though that approach is certainly possible, it also appears unnecessarily lengthy and cumbersome.

\subsection{Spherical space-time}

We are considering the problem of the dynamical generation/perturbation of a black hole in a model with spherical symmetry from/with a pure radiation field. To this end we need first to specify the explicit form of the field equations for a spherically symmetric space-time. For this purpose it is convenient to use a coordinate system $x^{\alpha}\equiv (x^0,x^1,\theta,\phi)$, with $x=\{x^0,x^1\}$ the coordinates of the two-spaces with ${(\theta,\phi)}=$constant, and to write the line element of the physical metric $g_{\mu\nu}$ as
\be \label{eq:lineg}
ds^2=g_{ab}(x)dx^a dx^b + r^2(x)d\Omega^2,
\en
and the line element of the auxiliary metric $h_{\mu\nu}$  as
\be \label{eq:lineh}
d\tilde{s}^2=h_{ab}(x)dx^a dx^b + \tilde{r}^2(x)d\Omega^2.
\en
In the above expressions, $d\Omega^2$ is the line element of a two-sphere. While these line elements are formally equal, their relation is non-trivial due to the relative deformation between the metrics defined by the matrix $\hat{\Sigma}$, which in general implies that $r^2(x)\neq \tilde{r}^2(x)$. Our strategy will be to solve the field equations (\ref{eq:fieldequationsdefinitiveX}) involving the metric $h_{\mu\nu}$, and then use the transformation matrix $\hat{\Sigma}$ to obtain  $g_{\mu\nu}$. In this sense, we note that using (\ref{eq:sigmafinal}) in (\ref{eq:h-g-relation}), the relation between the four-dimensional metrics $h_{\mu\nu}$ and $g_{\mu\nu}$ boils down to
\begin{equation}\label{eq:h-g}
g_{\mu\nu}= h_{\mu\nu}+2l_P^2\kappa^2\rho_{in} l_\mu l_\nu \ , \  g^{\mu\nu}= h^{\mu\nu}-2l_P^2\kappa^2\rho_{in} l^\mu l^\nu \ .
\end{equation}
Since we are considering a radially infalling fluid, $l_\mu=(l_0,l_1,0,0)$, the above relations guarantee that in this problem $r^2(x)=\tilde{r}^2(x)$ and
$g_{ab}(x)=h_{ab}(x)+2l_P^2\kappa^2\rho_{in} l_a l_b$, where $l_a=\{l_0,l_1\}$, which simplifies the analysis. The tilde on top of the function $r(x)$ will thus be omitted from now on.

Using the line element (\ref{eq:lineh}), the components of the Einstein tensor can be written as \cite{Poisson90}
\bea
^4G_{ab}(h)&=&-\frac{1}{r^2}[2r r_{;ab}+(1-2r\Box r-r^{,a}r_{,a})] \label{eq:G1} \\
G_{\theta \theta}(h)&=&\sin^2 \theta G_{\phi \phi}(h)=r\Box r -\frac{1}{2}r^2\bar{R}, \label{eq:G2}
\ena
where $\bar{R}$ and all covariant derivatives are computed using the two-dimensional  metric $h_{ab}$  (with latin indices to distinguish it from the four-dimensional metric $h_{\mu\nu}$). Note that these equations are completely general for any spherically symmetric metric, as we have not specified any coordinates associated to the hypersurfaces $(\theta,\phi)=$constant.

Combining the right-hand side of (\ref{eq:fieldequationsdefinitiveX}) with Eqs.(\ref{eq:G1}) and (\ref{eq:G2}),  we find
\bea
2rr;_{ab}+(1-2r\Box r-r^{,a}r_{,a})h_{ab}&=&-\kappa^2 r^2 T_{ab} \label{eq:G1b}\\
\Box r -\frac{1}{2}r \bar{R}&=& \kappa^2 r P \label{eq:G2b} \ ,
\ena
where $T_{ab}=\rho_{in} l_a l_b$ and $P=T_{\theta\theta}=T_{\phi\phi}=0$ for a presureless fluid.

It is now convenient to introduce the scalar functions $A=A(x)$ and $m=m(x)$, such that $A(x)\equiv r^{,a}r_{,a}=1-2m/r$, in terms of which (\ref{eq:G1b}) becomes
\be
r_{;ab}+\left(\frac{{m}}{r^2}-\Box r\right)h_{ab}=-\frac{ r \kappa^2}{2} T_{ab} \ . \label{eq:G1c}
\en
Taking the trace of this equation, we obtain $\Box r=\frac{2{m}}{r^2}+\frac{ r \kappa^2}{2} T$, and (\ref{eq:G1b}) is written as
\be \label{eq:G1d}
r_{;ab}-\frac{{m}}{r^2} h_{ab}=-\frac{ r \kappa^2}{2} (T_{ab}-h_{ab}T).
\en
Replacing now the $\Box r$ term obtained above back into (\ref{eq:G2b}), one obtains
\be \label{eq:G2c}
\bar{R}=\frac{4m}{r^3}+\kappa^2(T-2P) \ .
\en
Consider now the quantity $\partial_a A$, which establishes the following relation between $r_{;ab}$ and $m_{,a}$
\begin{equation}
m_{,a}=\frac{m}{r}r_{,a}-h^{bc}r r_{,b}r_{;ac} \ .
\end{equation}
Inserting (\ref{eq:G1d}) in this equation, we find
\be \label{eq:massderivative}
{m}_{,a}=\frac{\kappa^2r^2}{2} r_{,b}(T_a^b-\delta_a^b T).
\en

Eqs.(\ref{eq:G1d}), (\ref{eq:G2c}) and (\ref{eq:massderivative}) constitute the system of equations defining our problem. We already noted that the null fluid satisfies $T=0$, and imposed $P=0$ for simplicity.  The resulting field equations in such a scenario are
\bea
r_{;ab}-\frac{{m}}{r^2} h_{ab}&=&-4\pi r\rho_{in} l_a l_b \label{eq:G1e} \\
\bar{R}&=&\frac{4m}{r^3} \label{eq:G2e} \\
m_{,a}&=&4\pi r^2 T_{a}^{b}r_{,b}. \label{eq:G3d}
\ena

\section{Formation of black holes}\label{sec:IV}

In this section we solve Eqs.(\ref{eq:G1e}), (\ref{eq:G2e}) and (\ref{eq:G3d}) corresponding to an ingoing stream of particles represented by a presureless null fluid. We will thus focus on a Vaidya-type metric \cite{Vaidya70} of the form
\begin{equation}\label{eq:line-V1}
d\tilde{s}^2=-A e^{2\psi}dv^2+2e^{\psi} dvdr+r^2d\Omega^2 \ ,
\end{equation}
where $A=1-\frac{2m(v)}{r}$ and $\psi$ are functions of $x=\{v,r\}$, and the null vector  is normalized as $l_a=-\partial_a v$. This type of metric is often used in the study of black hole formation and evaporation processes \cite{NS-F}.

The $r-v$ component of (\ref{eq:G1e}) gives $\psi_{,r}=0$ and implies that $\psi=\psi(v)$. This means that we can define a new variable $V(v)$ such that $dV= e^{\psi(v)} dv$, and eliminate this function from the line element (\ref{eq:line-V1}). As a result, the only non-trivial function in the problem is $m=m(v,r)$. Using the $v-v$ equation of (\ref{eq:G1e}) or the $v-$component of (\ref{eq:G3d}), we find
\begin{equation}\label{eq:m_v}
m_{,v}=\frac{\kappa^2 r^2}{2}\rho_{in} \ .
\end{equation}
The $r-$component of (\ref{eq:G3d}) gives $m_{,r}=0$ and implies that $m=m(v)$, which can also be verified using the conservation equation of the matter. As a result, the function on the right-hand side of  (\ref{eq:m_v}) can only depend on the $v$ coordinate. We will thus introduce a function $L(v)=\frac{\kappa^2 r^2}{2}\rho_{in}$, called luminosity, to characterize the flux of incoming radiation, so that $m_{,v}=L(v)$ and $m(v)=\int^v L(v')dv'$.

In order to write the solution in terms of the physical metric $g_{\mu\nu}$, let us define the corresponding Vaidya-type metric as
\be \label{eq:EF-g}
ds^2=-Be^{2\Psi}dv^2+ 2e^{\Psi}dv dr+r^2 d\Omega^2
\en
where $B(v,r)=1-\frac{2M(v)}{r}$. Comparing with (\ref{eq:line-V1}), we find that $\Psi=0$ due to the $r-v$ component of (\ref{eq:G1e}). From Eq.(\ref{eq:h-g}), it follows that

\begin{equation}
A=B+2l_P^2 \kappa^2 \rho_{in} \ ,
\end{equation}
which implies
\be \label{eq:massrelation}
M(v)=m(v)+ r l_P^2 \kappa^2 \rho_{in}.
\en
Replacing this expression in the definition of $B$ and using the definition of the luminosity given above, we obtain
\be \label{eq:f-final}
B=1-\frac{2\int^{v}L(v')dv'}{r}-\frac{4L(v)}{\rho_P r^2} ,
\en
where $\rho_P=\frac{c^2}{l_P^2 G} \sim 10^{96} kg/m^3$ is Planck's density. This formula is the main result of this work. Let us analyze now its implications.

The space-time (\ref{eq:f-final})  resulting from the perturbation of a Schwarzschild black hole by a null fluid in the Palatini theory $f(R,Q)=R+l_P^{2}(a R^2 + Q)$  is formally that of a (nonrotating) charged (Reissner-Nordstr\"om) black hole but with a charge term whose sign is the opposite to the usual one. This allows us to define an ``effective charge" in (\ref{eq:f-final}) as
\be\label{eq:eff_charge}
Q^2(v)=\frac{4L(v)}{\rho_P},
\en
and a  mass term
\begin{equation}\label{eq:eff_mass}
\tilde{M}(v)=\int_{v_0}^{v}L(v')dv'  ,
\end{equation}
such that (\ref{eq:f-final}) becomes
\begin{equation}\label{eq:f-finalX}
B=1-\frac{2\tilde{M}(v)}{r}-\frac{Q^2(v)}{r^2} \ .
\end{equation}
The horizons of this ``opposite" sign charged metric, solutions of $r^2-2\tilde{M}(v)r-Q^2(v)=0$, are found at
\be \label{eq:horizons}
r_{\pm}=\tilde{M}(v) \pm \sqrt{\tilde{M}(v)+Q^2(v)}.
\en
Since $r_-<0$ it follows that there is a single horizon, $r_+$, larger than the usual  Schwarzschild radius, i.e., $r_+>2\tilde{M}$.
\begin{figure}[h]
\includegraphics[width=0.45\textwidth]{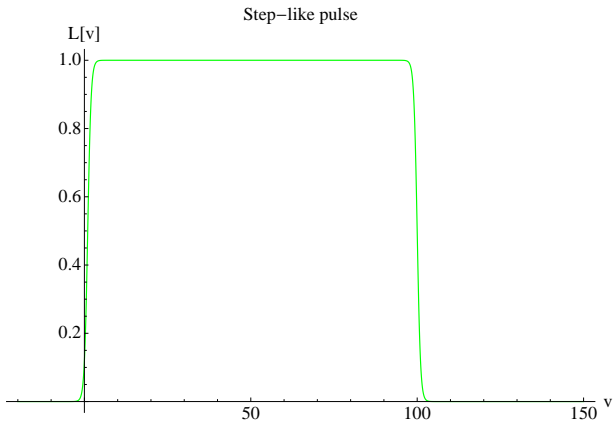}
\caption{Smooth ingoing flux (\ref{eq:pulse})  mimicking a step-like profile to illustrate the formation of a Schwarzschild black hole out of Minkowski space.}
\label{fig:1}
\end{figure}

\begin{figure}[h]
\includegraphics[width=0.45\textwidth]{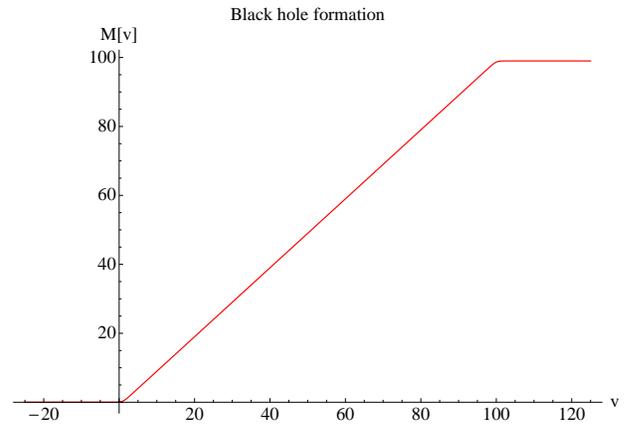}
\caption{Behaviour of the black hole mass function $\tilde{M}(v)$ due to the step-like pulse (\ref{eq:pulse}).}
\label{fig:2}
\end{figure}

The dynamics of these ``charged" black holes is strongly tied to the shape and temporal distribution of the ingoing matter stream. In this sense, it must be noted that a typical Dirac-delta profile for the luminosity function, i.e. $L(v)=m_0 \delta(v-v_0)$ is not consistent because that would imply a divergent metric along the path $v=v_0$ due to the charge term (\ref{eq:eff_charge}).  This property of the metric forces the consideration of smooth and finite luminosity profiles. Moreover, in order to guarantee a smooth geometry, the luminosity function should be continuous and differentiable at least up to order two. As an example, let us consider the following step-like radiation pulse (see Fig.\ref{fig:1}):
\be \label{eq:pulse}
L(v) = \frac{\tilde{M}_0}{4(v_1-v_0)}(1+\tanh(v-v_0))(1-\tanh(v-v_1)),
\en
with $v_0=1$ and $v_1=100$ in Fig.\ref{fig:1} signaling the (effective) beginning and end of the pulse, respectively. This flux of radiation has the following effect on the geometry. We begin with an empty Minkowski space at $v\to -\infty$. Near $v\approx v_0$, the amplitude of the flux grows up to its maximum and remains almost constant until $v\approx v_1$.  During the time that the pulse is active, the mass (\ref{eq:eff_mass}) grows linearly (see Fig.\ref{fig:2}), and the effective charge remains constant. Once the radiation pulse is off (beyond $v_1$), the charge disappears and the configuration becomes that of a Schwarzschild black hole with constant mass $\tilde{M}\approx \tilde{M}_0$.\\

\begin{figure}[h]
\includegraphics[width=0.45\textwidth]{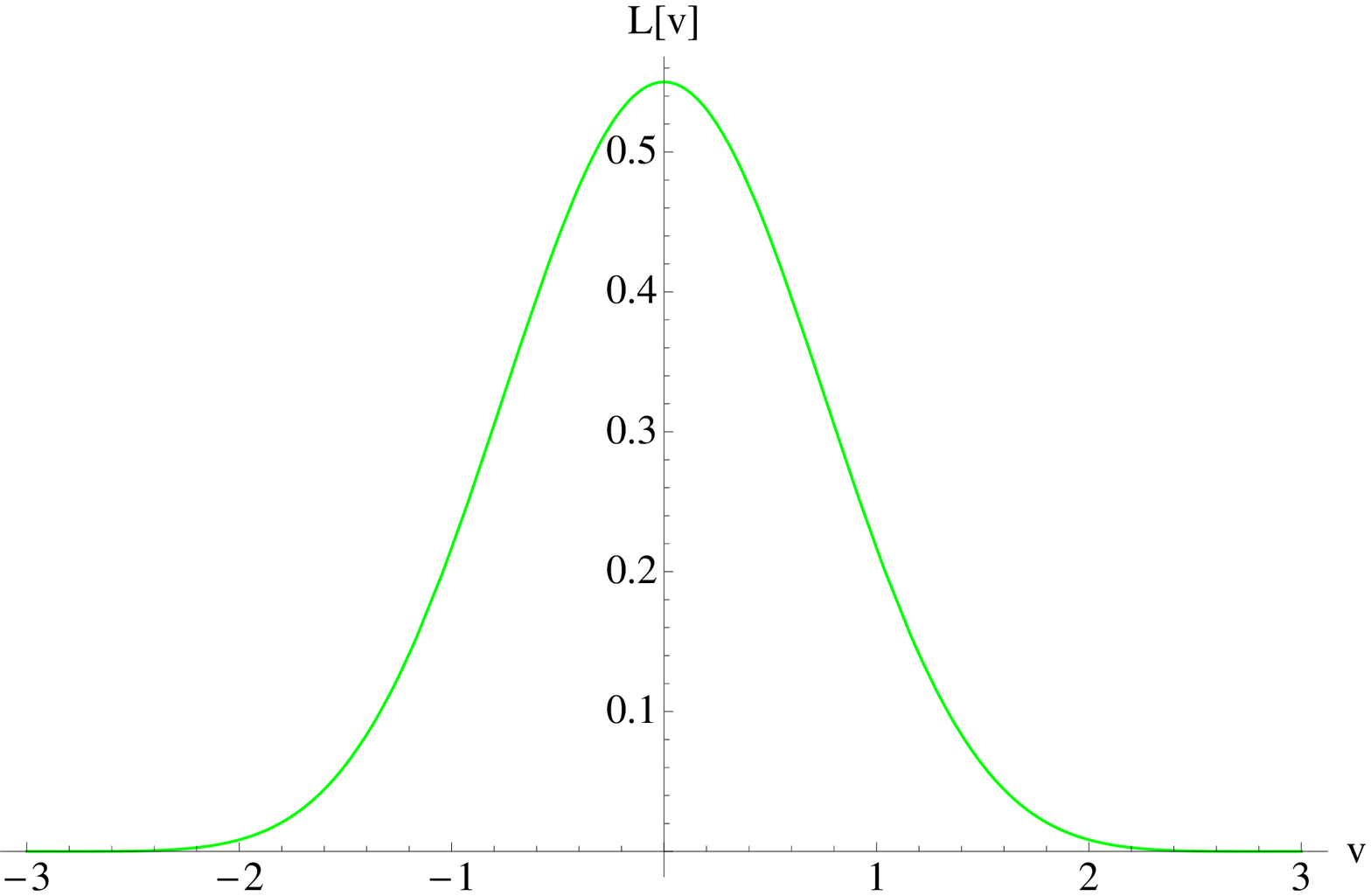}
\caption{$L(\upsilon)$ profile (\ref{eq:wavelet}) for a compact support perturbation. }
\label{fig:3}
\end{figure}

\begin{figure}[h]
\includegraphics[width=0.45\textwidth]{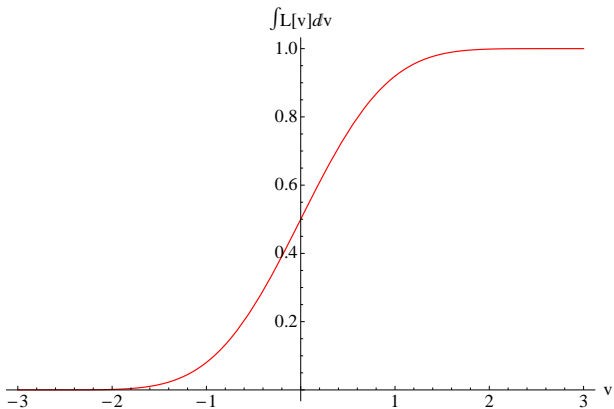}
\caption{Growth of the black hole mass function $\tilde{M}(v)$ due to the perturbation (\ref{eq:wavelet}).}
\label{fig:4}
\end{figure}

Consider now a Schwarzschild black hole of mass $\tilde{M}=\tilde{M}_0 $ and a compact support perturbation of the form (see Fig.\ref{fig:3})
\begin{eqnarray}\label{eq:wavelet}
L(v)&=&\frac{\epsilon}{240} \left( |v-3|^5-6  |v-2|^5+15 |v-1|^5\right. \nonumber \\&-& \left. 20|v|^5+15 |v+1|^5-6 |v+2|^5+ |v+3|^5\right) .
\end{eqnarray}
This function and its derivatives  are continuous and differentiable up to third order and are exactly zero beyond $|v|\ge 3$ (unlike the pulse (\ref{eq:pulse}), which is defined over the whole real line). This flux provides a local perturbation of an existing Schwarzschild black hole and complements the formation scenario of the previous example. The integral $\int_{-3}^3 L(v)$ is normalized to unity and, therefore, $\tilde{M}(v)=\tilde{M}_0+\epsilon$ for $v\ge 3$ (see Fig.\ref{fig:4}).

\section{Conclusions and perspectives} \label{sec:V}

In this paper we have considered the formation and (exact) perturbation of spherically symmetric black holes in a Palatini $f(R,Q)$ theory by means of neutral fluxes of a presureless null fluid. We have seen that once the flux of radiation is over, the resulting geometry is exactly the Schwarzschild solution \cite{footnote} of GR.  Though the asymptotic solution is exactly Schwarzschild, the intermediate perturbed states are not. As long as the ingoing flux is non-zero, the metric exhibits a charge-like term whose form is directly related to the luminosity function of the radiation stream. Though this term is Planck suppressed, its mere existence indicates that the energy density of the matter is leaving a direct imprint on the metric, putting forward that any fluctuation of the matter profile will induce an identical fluctuation on the space-time metric modulated by a factor  $1/r^2$, as given in Eq.(\ref{eq:f-final}). This shows, with an exact analytical solution, that in these Palatini theories the backreaction of the matter on the geometry is completely sensitive to the most minute details of the matter profile, $Q^2(v)$, not just to its integrated form, $\tilde{M}$. This, in particular, motivates the consideration of the process of black hole evaporation via Hawking radiation in these scenarios to see whether the imprint that the infalling matter leaves on the geometry can be recovered at infinity through the interaction of the geometry with the outgoing Hawking quanta. This point naturally opens the door to the study of the  interaction between the ingoing $\rho_{in}$ and outgoing $\rho_{out}$ fluxes to see whether a kind of Dray-'t Hooft-Redmount (DTR) \cite{DTR} relation may be found in our model. A difficulty that may arise in our case in comparison with GR is that the peculiarities of our theory prevent the use of delta functions to describe the matter profile. In addition, the consideration of two (ingoing and outgoing) fluxes implies that the curvature scalar  $Q$ becomes a function of the product $\rho_{in}\rho_{out}$ \cite{JGD2012}. The absence of that kind of products on the right-hand side of Einstein's equations was crucial in the DTR analysis and also in the study of the mass inflation phenomenon \cite{Poisson90} (see also \cite{massinflation}) in the charged case. These aspects will be studied in detail and reported elsewhere. \\

\acknowledgments

The work of G. J. O. has been supported by the Spanish grant FIS2008-06078-C03-02, FIS2011-29813-C02-02,  and the JAE-doc program of the Spanish Research Council (CSIC). D. R. -G. thanks the hospitality of the theoretical physics group at Valencia U., where part of this work was carried out.


\begin{thebibliography}{99}
\bibitem{footnote}
From our derivations, it is straightforward to verify that the addition of a cosmological constant or the consideration of an $\tilde{f}(R)$ piece with non-quadratic corrections would lead to Schwarzschild-de Sitter solutions.
\end{thebibliography}
\end{document}